\documentclass[conference]{IEEEtran}

\usepackage{enumerate}
\usepackage{epsfig}
\usepackage{graphicx}
\usepackage{epstopdf}
\usepackage{amsmath}
\usepackage[linesnumbered,ruled]{algorithm2e}
\usepackage{xurl}
\usepackage{listings}
\usepackage{subfigure}
\usepackage{rotating}
\usepackage{lscape}
\usepackage{verbatim}
\usepackage{hhline}
\usepackage{textcomp,booktabs}
\usepackage{xcolor}
\usepackage{paralist}
\usepackage{cite}
\usepackage{amsthm}
\usepackage{flushend}
\usepackage{hyphenat}
\usepackage{enumitem}
\usepackage{multicol}
\usepackage{color, colortbl}
\usepackage[autostyle,threshold=0]{csquotes}
\usepackage{stfloats}

\newcommand{\newmaterial}[1]{\textcolor{black}{#1}}

\begin{document}

\title{Victim-Centred Abuse Investigations and Defenses for Social Media Platforms}

\author{\IEEEauthorblockN{Zaid Hakami\IEEEauthorrefmark{1},\IEEEauthorrefmark{2},
Ashfaq Ali Shafin\IEEEauthorrefmark{1},
Peter J. Clarke\IEEEauthorrefmark{1}, 
Niki Pissinou\IEEEauthorrefmark{1} and
Bogdan Carbunar\IEEEauthorrefmark{1}}
\IEEEauthorblockA{\IEEEauthorrefmark{1}Florida International University, USA}
\IEEEauthorblockA{\IEEEauthorrefmark{2}Jazan University, Saudi Arabia}
\IEEEauthorblockA{\IEEEauthorrefmark{1}\{zhaka001, ashaf016, clarkep, pissinou, carbunar\}@fiu.edu}}

\IEEEoverridecommandlockouts
\makeatletter\def\@IEEEpubidpullup{6.5\baselineskip}\makeatother
\IEEEpubid{\parbox{\columnwidth}{
		Symposium on Usable Security and Privacy (USEC) 2025 \\
		24 February 2025, San Diego, CA, USA \\
		ISBN 979-8-9919276-5-9 \\
		https://dx.doi.org/10.14722/usec.2025.23020 \\
		www.ndss-symposium.org, https://www.usablesecurity.net/USEC/
}
\hspace{\columnsep}\makebox[\columnwidth]{}}

\maketitle

\pagestyle{plain}

\begin{abstract}
Online abuse, a persistent aspect of social platform interactions, impacts user well-being and exposes flaws in platform designs that include insufficient detection efforts and inadequate victim protection measures. Ensuring safety in platform interactions requires the integration of victim perspectives in the design of abuse detection and response systems. In this paper, we conduct surveys (n = 230) and semi-structured interviews (n = 15) with students at a minority-serving institution in the US, to explore their experiences with abuse on a variety of social platforms, their defense strategies, and their recommendations for social platforms to improve abuse responses. We build on study findings to propose design requirements for abuse defense systems and discuss the role of privacy, anonymity, and abuse attribution requirements in their implementation. We introduce ARI, a blueprint for a unified, transparent, and personalized abuse response system for social platforms that sustainably detects abuse by leveraging the expertise of platform users, incentivized with proceeds obtained from abusers.

\end{abstract}

\section{Introduction}


Social platform affordances that enable users to publish personal content, establish relations and share information with contacts, also help them maintain and increase social capital~\cite{ESL07, BKM11}, and facilitate organizational efforts~\cite{SP12}. However, the platform promises to provide safe spaces of expression, and factors like anonymity and implicit and difficult-to-detect abuse~\cite{WRE21,CBMKG20} make it challenging to moderate user activities. This leads to many users experiencing abuse on social platforms and to criticism of platform responses to abuse~\cite{WK22, SCP23, PewResearch2021, D17, TABBBCDDKK21, LYZP16}, that reveal the failure of social platform designs to consider the malicious use of their features to abuse users.

This suggests the need to re-design platform interactions and abuse defenses to focus on the well-being of users. While this requires factoring in insights from past victims, we observe a lack of understanding of (1) the abuse defenses adopted by victims and their perceptions about the defenses that platforms should adopt, and (2) their relation to the impact of abuse experienced by victims, captured in the context of the diverse types of abuse they encountered on the multiple social platforms they use. In this paper, we investigate the perspectives of social platform abuse victims on platform defense design by focusing on the following research questions:

\begin{compactitem}

\item
{\bf RQ1}: 
What kinds of abuse do users experience on social platforms? (a) Where do they experience the most abuse, and what is its impact?

\item
{\bf RQ2}:
How do abuse victims respond to abuse? (a) Does the impact of abuse they experience affect the defenses they adopt?

\item
{\bf RQ3}:
How do abuse victims want social platforms to address the abuse they experienced? (a) Are their preferences related to the impact of experienced abuse?

\end{compactitem}

\noindent
To investigate these questions, we build on literature findings that young, minority users are more likely to be impacted by online abuse~\cite{ES19, GVF16, KBX16, KMDL+16, MGC23, SBDDGWKTA23, ISIGWBAFGGN22, FF22, UCW21}, and on the conjectures that such users are more likely to have developed strategies to avoid abuse and respond to encountered abuse. More specifically, we use a mixed-methods study to investigate the experiences of students from a minority-serving institution in the US. The study consists first of a survey with ($n$ = 230) recruited students, that uses a 1-to-5 abuse impact score we introduce, to probe exposure to online abuse that includes threats, blackmail, doxxing, harassment, hate speech, and abusive messages in social platforms that include social networks (Facebook, Instagram), social media (Twitter/X, TikTok), and messaging apps (WhatsApp, Snapchat). Second, the study consists of interviews conducted with a subset ($n$ = 15) of the survey respondents in the context of their reported exposure to at least one instance of severe abuse.

Our study reveals that young and educated minority users are exposed to substantial, often highly impactful abuse on a variety of social platforms, some of which emerge to be more conducive toward abuse. We observe cross-pollination of abuse across multiple platforms, where abusers leverage the affordances of diverse platforms to expand their knowledge of victims and amplify their abuse attacks (RQ1). We find a statistically significant correlation between the impact of abuse experienced by survey respondents and their adoption of defenses against abusers, where victims of high-impact abuse are more likely to adopt each of the investigated defenses than low-impact victims (RQ2). In addition, while some interview participants mastered the abuse defense arsenal provided by social platforms, including using pseudonymous accounts to protect their identity from abusers and to sandbox their activities on different accounts and platforms, other participants experienced heightened anxiety following abuse from anonymous contacts (RQ1 and RQ2). This reveals that even regular users driven by their emotions and not by financial incentives (e.g., hackers or influence operatives) can thwart social platform defenses to create fake, sockpuppet accounts that are sufficiently complex to push victims into repeat abuse. \newmaterial{Further, while participants revealed diverse preferences for social platform responses to abuse, we found no significant association between the impact of abuse experienced and preferences for defenses (RQ3).}

The ability of users to connect to and abuse other users suggests intrinsic flaws in the design of social platforms and antiquated abuse response protocols, e.g., allowing users to create multiple pseudonymous accounts even after being involved in abuse and failing to detect abuse and take appropriate action, even after victims report the abuse. To address these issues, we first crystallize study findings into design requirements for abuse defense systems. We then leverage these requirements to heed community calls for more defense options that integrate the needs of vulnerable users~\cite{SDBHACRAMA21,BDSL17,BBV18}.

More specifically, we introduce ARI, a blueprint for unified, transparent, and personalized social platform abuse defense systems. ARI seeks to establish mutually beneficial collaborations between victims and platforms. ARI-compliant platforms provide prompt, transparent services to abuse victims, including issuing certificates of abuse, alerting them when they become vulnerable to abuse from multiple accounts controlled by the same person, and notifying them of progress during the abuse verification process. Conversely, platforms leverage the expertise of their users to crowdsource the abuse verification process to qualified verifiers, and collect evidence about the activities of abusers while charging the cost of implementing this process to verified abusers and abusive reporters.

\section{Background and Related Work}
\label{sec:related}

\noindent
{\bf Online Abuse and its Impact on Women and Minority Groups}.
With the proliferation of mobile devices and the ubiquity of Internet access for users in most regions of the world, online abuse is becoming a growing issue~\cite{PewResearch2021, TABBBCDDKK21}. Studies show that nearly half of all Internet users have encountered some form of online abuse, with the prevalence varying significantly by country~\cite{TABBBCDDKK21}. In the United States, between 40\% and 50\% of adult internet users report having experienced online abuse ~\cite{PewResearch2021, LYZP16}. The term {\it online abuse} encompasses a wide range of harmful behaviors that include {\it hate and abusive speech, threats, doxxing, blackmail, harassment, stalking, impersonation, and public shaming} ~\cite{BDSL17, LYZP16}.

Online abuse is often based on identity~\cite{HU24, KCN23, C-GCC23, Citron22, FF22, GPV22, ISIGWBAFGGN22, NDS+21, RMCS+20, Ruv21, UCW21, VCSA17}. For instance, through surveys with users from 14 world regions, Im et al.~\cite{ISIGWBAFGGN22} show that women perceive greater harm from online harassment than men and tend to prefer platform responses to compensatory responses to abuse. The survey of more than 16K respondents in 10 countries by Henry and Umbach~\cite{HU24} about the prevalence of victimization and perpetration of sextortion reveals that LGBTQ+, men, and younger respondents are more likely to report sextortion victimization and perpetration. Batool et al.~\cite{BNT24} reveal that in Pakistan’s patriarchal, honor-based society, even non-sexual images and manipulated images are used for blackmail and sextortion. Musgrave et al.~\cite{TAS22} document the harassment experienced online by Black women and femmes in the US, while Francisco and Felmlee~\cite{FF22} reveal the targeted nature of the harassment encountered by Hispanic/Latinx and Black women through tweets.

The current work builds on previous observations that addressing online abuse requires an understanding of the experience of the victims and how they want it to be addressed, given Im et al.~\cite{ISIGWBAFGGN22}'s insight that those who design and run social platforms are unlikely to have faced online abuse themselves to the level described in the works previously mentioned. We observe, however, limitations in our understanding of the exposure of young, educated users to abuse experienced on the range of various social platforms they currently use. We conjecture that tech-savvy users have developed effective strategies to avoid and respond to encountered abuse and thus have the potential to provide insights that can help improve platform abuse defense. To address these limitations and explore our conjecture, we survey students at a minority institution about their experiences with six types of abuse, the social platforms on which they experience it, and its impact (RQ1). In addition, we seek deeper insights through a qualitative study with participants exposed to more severe forms of abuse, conducted in the context of specific instances of abuse they experienced.

\noindent
{\bf Social Platform Abuse Responses and Perceptions}.
Online abuse directly impacts not only users but also social media platforms. This is because reactions to abuse that include closing accounts, reduced user activities, and smaller social circles can reduce revenues, e.g., from ads, for social platforms~\cite{PewResearch2021,D17,LYZP16}. While 84\% of U.S. social media users believe that these platforms should protect them against harassment \cite{ADL2019OnlineHate}, recent years have witnessed a decline in the public's confidence in the ability of platforms to regulate online conduct~\cite{VergeTechSurvey2020, BDSL17, TAS22, SSHCS21}. Indeed, while most social platforms have policies and moderation processes to remove content that violates community guidelines, e.g.,~\cite{G20, G18}, they often fail to eradicate abuse on their platforms~\cite{PewResearch2021,D17}.

Content moderation involves both manual and automatic detection. Some platforms rely on human workers to identify content that violates policies~\cite{G20,G18}. This work is often done by volunteer moderators~\cite{W19} or underpaid workers who must make quick decisions with little context, often involving traumatizing content~\cite{G20}. However, users still consider the process to be inefficient~\cite{W17}. 

Using a combination of AI/ML and NLP tools to identify abuse is a promising direction~\cite{CSSG17, BDSL17, APP19, ZRT18, MBD23, TwitterModeration}, that however suffers from incorrect classification issues~\cite{APP19}, bias~\cite{B21, BYSB21}, and also fails to address structural power imbalances~\cite{BDSL17}. Extreme defenses like deplatforming abusers have also been shown to fail, resulting in only temporary disruptions followed by abusers reconvening on alternate platforms and channels~\cite{VHA24}.

Redmiles et al.~\cite{RBB19} reveal that perceptions of digital privacy, security, and community values shape online safety. Wilkinson and Knijnenburg~\cite{WK22} find that heightened vulnerability and severity of experienced online abuse enhance safety protection behaviors in Caribbean communities. Im et al.~\cite{ISIGWBAFGGN22} explore perceptions of not only standard platform responses to abuse, e.g., abuser account suspensions and content moderation, but also public shaming and restorative justice, e.g., apologies and victim compensation. They report that women from 14 regions tend to prefer existing platform responses to compensation. However, Schoenebeck et al.~\cite{SHN21} highlight that content moderation practices primarily focus on punishing abusers~\cite{SCP23}, insufficiently supporting harassment victims.

Regarding responses to reported abuse, Schoenebeck et al.~\cite{SSHCS21} find that many users are skeptical about the fairness of platform resolutions. Musgrave et al.~\cite{TAS22} show that studied Black women and femmes often avoid reporting instances of gendered and racist harassment on platforms, believing that reporting would not be beneficial. Sambasivan et al.~\cite{sambasivan2019they} also note hesitance among South Asian women to use platform reporting features, often due to the platform's inability to comprehend the context of regional-specific issues. In the case of sextortion, Wolak and Finkelhor\cite{wolak2016sextortion} reveal that the severity of the abuse leads many female teenager and young adult victims to hesitate before reporting the abuse due to shame and doubts about the effectiveness of reports. Ahmed~\cite{ahmed2021complaint}'s notion of {\it strategic inefficiency}, where institutions deliberately delay handling complaints, can partially explain the perceived ineffectiveness of social platform abuse reports.

Our study investigates the responses of young and educated victims to abuse experienced on various social platforms and their perceptions of existing platform defenses and their effectiveness (RQ2). We build on Uttarapong et al.~\cite{UCW21}'s finding that anonymous viewers have an advantageous position on privacy vs. exposed and vulnerable streamers to investigate perceptions of privacy and anonymity features provided by platform affordances from the perspectives of both regular users and abuse victims.

\noindent
{\bf Design of Next Generation Abuse Responses}.
Meta has recently adopted account verification, promising subscribers proactive account protection against impersonation attacks and direct support from a real person~\cite{R23}. However, the associated monthly fees, e.g., \$12 for Facebook, ensure that this service is outside the reach of an overwhelming majority of platform users and suggest a medieval frame of mind for addressing abuse. Instead, users' perceived limitations of existing social platform defenses suggest the need for improved moderation interfaces to protect victims from abuse. Previous studies of diverse forms of abuse recommend that the design of social platform defenses should move beyond blocking abusers and deactivating accounts~\cite{BDSL17, BBV18, RBC20, SHN21, VCSA17}. Schoenebeck et al.~\cite{SCP23} reveal the need for defenses to move beyond one-size-fits-all approaches and consider more victim-centered strategies in combating online abuse. Blackwell et al.~\cite{BDSL17} argue that fully addressing abuse requires the integration of the needs of vulnerable users into the design and moderation of online platforms.

Previous work argues that the evolving nature of content moderation requires a shift toward models that promote procedural justice and holistic transformations. Katsaros et al.~\cite{KKT24} advocate for integrating principles of fairness, transparency, and respect into moderation practices to reduce violations and enhance platform legitimacy. Vitak et al.~\cite{VCSA17} argue for tailored interventions to improve the safety and well-being of women on social media platforms. Similarly, Xiao et al.~\cite{XCS22} call for systemic changes to effectively address online harms among adolescents. Innovative solutions are emerging from various studies proposing specific tools and systems to support victims. Sultana et al.~\cite{SDBHACRAMA21} designed a system to help women collect evidence of harassment from Facebook Messenger and share it with social contacts to prove authenticity. Goyal et al.~\cite{GPV22} propose a prototype tool to assist female journalists in managing online harassment, focusing on crisis management and recovery steps.

While diverse and user-tailored abuse defense mechanisms are of paramount importance, Wei et al.~\cite{WCKKRT23} show that even experts disagree on the prioritization of abuse types and mitigation advice. This poses significant challenges in developing effective defense tools and highlights the crucial need for adaptable and inclusive approaches to combat online abuse. Wei et al.~\cite{WCKKRT23}'s study suggests the need for a better understanding of user perceptions of the missing components required to improve their responses to experienced abuse, which is one of the focus points of this research (RQ3). We use findings from our study to identify abuse defense design requirements and use them to introduce ARI, a blueprint for defense solutions that leverage cryptography, digital snapshots, and abuser de-anonymization techniques to provide a robust support system that encourages collaboration between platforms and victims.

Particularly relevant to this paper is the work of Kim et al.~\cite{KLJK24}, who use design workshops to understand the functionality users need to protect themselves from abuse when posting content on X-like social platforms. Kim et al.~\cite{KLJK24} leverage the workshops to introduce design goals,

and use them to develop Re:SPect, a concept tool that allows users to set post-level access controls, provides a summarization of responses received for a post, and enables users to respond to clusters of comments. Our study crystallizes a set of abuse defense system requirements that apply to a variety of social platforms, including social media like Twitter/X, social networks like Facebook and Instagram, and messaging apps like WhatsApp and Snapchat. ARI, the blueprint defense system we propose, includes more general functionality than Kim et al.~\cite{KLJK24}'s Re:SPect, e.g., allowing users to document, respond to, and report abuse, and provides victims with cues that their abuse experiences improve platform operations.

\section{Methods}

To explore the impact of abuse on social platform users and their responses, we designed a mixed-methods study consisting of an online survey and semi-structured interviews. This section describes participant recruitment, the survey and interview design, and data analysis.

\subsection{Participant Recruitment}
\label{sec:methods:demographics}

Survey participants were recruited through emails and posts in WhatsApp groups sent to local students. The emails included the survey link, while the social media posts contained a QR code for the link. At the end of the survey, participants were asked if they would like to participate in a one-on-one interview and provided their preferred contact information. Survey respondents who expressed interest, provided contact information, and experienced online abuse with at least an impact rating of 3, which they were willing to discuss, were invited to the interview. All interview participants were asked to distribute the survey link to their contacts. Survey participants were not paid. Interview participants were compensated 20 USD for their time. 

\subsection{Survey Design}
\label{sec:methods:survey}

\begin{table}[t]
\caption{Demographics of the 230 survey respondents, all recruited from minority-serving institution in the US.}
\textsf{
\scalebox{0.9}{
\begin{tabular}{llcc} 
\toprule
Demographic & Group & N & \% \\
\midrule
Gender & Man & 174 & 75.7\% \\
       & Woman & 47 & 20.4\% \\
       & Transgender & 3 & 1.3\% \\
       & Agender & 1 & 0.4\% \\
       & Prefer not to answer & 5 & 2.2\% \\
\midrule
Age & 18--25 & 168 & 73.0\% \\
    & 26--35 & 53 & 23.0\% \\
    & 36--45 & 2 & 0.9\% \\
    & Above 45 & 4 & 1.7\% \\
    & Prefer not to answer & 3 & 1.3\% \\
\midrule    
Race or Ethnicity & Hispanic, Latino, or Spanish origin & 70 & 30.4\% \\
                  & Hispanic and White & 48 & 20.9\% \\
                  & Asian alone & 55 & 23.9\% \\
                  & Black or African American alone & 23 & 10.0\% \\
                  & White alone & 24 & 10.4\% \\
                  & Middle Eastern or North African & 3 & 1.3\% \\
                  & American Indian or Alaska Native & 2 & 0.9\% \\
                  & Prefer not to answer & 5 & 2.2\% \\
\bottomrule
\end{tabular}}
}
\label{tab:survey:demographics}
\end{table}

The survey was designed by the research team over approximately three months. In the first step of the design phase, the team explored different types of online abuse and their impact, popular social platforms, and adopted defenses. The abuse types were selected by reviewing existing academic literature~\cite{TABBBCDDKK21, SBDDGWKTA23, WCKKRT23, WK22, PewResearch2021}, to capture perceptions about diverse abuse experiences with different severity levels. Our selection process was guided by two main criteria: the prevalence of each abuse type in prior research and the diversity of abuse experiences. In the second step, a focus group was conducted with four local K-12 teachers with expertise on online abuse acquired both from personal experience and from their students. The goal of the focus group was to refine the types of abuse and social platforms considered, as well as the survey questions identified in the first step. To minimize participant fatigue, the resulting survey focuses on six types of abuse (see Appendix~\ref{appendix:survey}): {\bf Threats} defined as expressions of intentions to inflict injury or damage; {\bf abusive messages}, encompassing inappropriate, offensive, or insulting communications that cause hurt or anger; {\bf doxxing}, the act of searching for and publishing private information to cause harm; {\bf blackmail}, involving coercion to prevent the distribution of sensitive private information; {\bf harassment}, characterized by unwelcome conduct that creates a hostile or unpleasant environment; and {\bf hate speech}, which includes expressions promoting hatred, violence, or discrimination.

\noindent
The first part of the survey investigates participants use of social platforms, including the ones they use the most. The second part explores participant experiences with the six types of abuse. For each type of experienced abuse, the survey further explores the social platform where the abuse occurred, strategies employed to address the abuse, and the impact of the worst instance of that type of abuse. The impact options are ranked from 1 to 5, where 1 denotes ``It did not affect me at all'', 2 - ``It affected me a bit'', 3 - ``It affected me but I was able to move on'', 4 - ``I was very affected and it was hard to move on'', and 5 - ``I was irreparably affected and am still feeling the effects''. The survey then asks participants about changes to social platforms that would help them better cope with abuse. Finally, the survey asks participants about their interest in a follow-up interview and their demographics.

\noindent
{\bf Data Analysis}.
We use the scipy.stats~\cite{SciPy} Python library to analyze the survey data. We use descriptive analyses and inferential statistical techniques. For instance, we use Chi-square tests to explore relationships between categorical variables and Spearman's rank correlation to assess monotonic relationships, offering insights into behavior and abuse impacts. For non-normally distributed data, we use Mann-Whitney U and Kruskal-Wallis tests to analyze ordinal data and groups with unequal variance. We apply post-hoc analyses with Bonferroni correction following significant test outcomes to mitigate Type I errors due to multiple comparisons.

\begin{table*}[h]
\centering
\resizebox{0.97\textwidth}{!}{
\textsf{
\begin{tabular}{lclcllcl}
\toprule
Participant & Gender & Race & Age & Student Type & Type of Abuse & Impact & Social Platform(s)\\
\midrule
P1 & M & Black or African American & 24 & Undergraduate & Doxxing/Hate Speech & 5 & Instagram, Facebook, Xbox\\
P2 & M & Hispanic, Latino, or Spanish origin & 25 & Graduate & Harassment & 4 & Discord \\
P3 & M & Hispanic, Latino, or Spanish origin, White & 24 & Undergraduate & Impersonation & 3 & Facebook \\
P4 & W & Asian & 25 & Graduate & Harassment & 4 & Snapchat \\
P5 & M & Hispanic, Latino, or Spanish origin & 25 & Graduate & Doxxing/Blackmail  & 3 & Facebook, WhatsApp\\
P6 & W & Asian & 30 & Graduate & Hate Speech & 5 & Slack\\
P7 & M & Asian & 25 & Graduate & Hate Speech & 3 & Facebook\\
P8 & W & Black or African American & 21 & Undergraduate & Hate Speech & 3 & TikTok\\
P9 & M & Asian & 29 & Graduate & Threats & 4 & Facebook\\
P10 & M & White & 21 & Undergraduate & Threats/Blackmail & 3 & Discord, Online Games\\
P11 & W & Hispanic, Latino, or Spanish origin & 21 & Undergraduate & Threats/Stalking & 5 & WhatsApp, Snapchat, Instagram\\
P12 & W & Asian & 25 & Graduate & Harassment & 3 & Instagram\\
P13 & M & Hispanic, Latino, or Spanish origin & 21 & Undergraduate & Harassment & 3 & Twitter/X, Valorant (video game). \\
P14 & W & White & 23 & Undergraduate & Threats/Stalking & 5 & Instagram\\
P15 & M & Hispanic, Latino, or Spanish origin & 22 & Undergraduate & Harassment/ Abusive Message & 3 & Discord \\
\bottomrule
\vspace{-5pt}
\end{tabular}}}
\caption{Demographics and abuse discussed by interview participants. Participants are balanced on gender and study background, and all experienced abuse with impact of at least 3 (it affected me but I was able to move on).}
\vspace{-14pt}
\label{tab:interview:demographics}
\end{table*}

\begin{figure*}
\centering
\subfigure[]
{\label{fig:socialnetsAll}{\includegraphics[width=0.85\columnwidth]{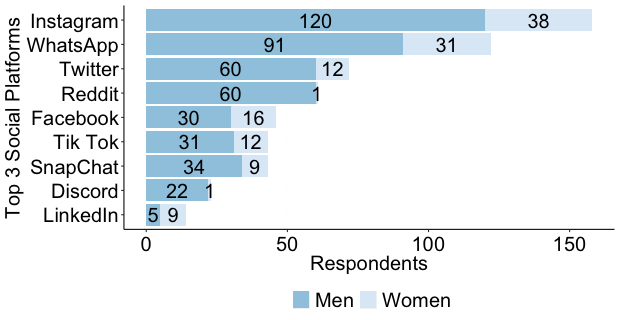}}}
\hspace{25pt}
\subfigure[]
{\label{fig:sn:mostabuse}{\includegraphics[width=0.85\columnwidth]{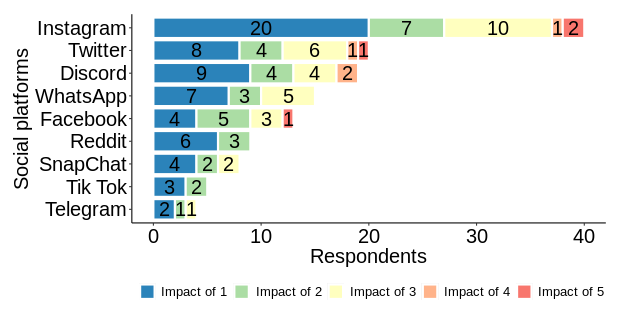}}}
\vspace{-7pt}
\caption{(a) Gender-based distribution of the social platforms rated top-3 most used by survey respondents.
(b) Platforms where survey respondents were exposed to most abusive messages. Discord is the least frequently mentioned top-3 platform but ranks third on user exposure to abusive messages. Respondents experienced abusive messages with impact above 3 on four platforms.}
\vspace{-14pt}
\end{figure*}

\subsection{Semi-Structured Interview Design}
\label{sec:methods:interview}

The semi-structured interview guide is designed to capture insights about participant experiences with abuse and their interactions with the social platforms where they experienced it. Each interview is conducted in the context of the most impactful instance of abuse experienced by the participant, which the participant feels comfortable discussing. The interview is structured into two parts (see Appendix~\ref{appendix:interview}). The first part identifies an abuse incident of significant impact that the participant is open to discuss, and explores its emotional and psychological impact, participant reactions, and adopted protective measures. The second part focuses on the role of social platforms in addressing abuse incidents and studies perceptions of platform responses, the support provided to victims, and their effectiveness in protecting from abuse. The interview further investigates suggestions for how social platforms should address abusive behaviors.

The interviews were conducted in person, in English, by one of the authors. Each session was audio-recorded with the participant's permission. The duration of the interviews ranged from 27 to 73 minutes ($M = 48$, $SD = 14.68$).

\noindent
{\bf Analysis Process}.
Interview responses were transcribed, pseudonymized, and securely stored. Transcripts were then analyzed using applied thematic analysis~\cite{GME11,GME12}, by systematically generating and iteratively conceptualizing codes and themes. Two researchers independently read the first three transcripts, coded responses to each interview question, and then organized them into themes through an initial codebook. The researchers met to discuss the themes and codes and revise the codebook. The researchers repeated this process over batches of 2-3 interview transcripts. After each batch, agreement was reached between the codes and themes identified by the researchers. After this process, the researchers independently applied the identified codebook to all the interview transcripts.

\subsection{Ethical Considerations}
\label{sec:Ethical}

The study procedure was scrutinized, and the full study was approved by our university's institutional review board. The consent form link was sent, and consent was obtained both before the survey and before the interview. During recruitment and the interview, we clearly declared the identity of the researchers, the research objective, the data we collect and how we process it, and the potential impact on the participant. Participants were explained the risks implied by participation, i.e., potential anxiety induced by remembering past abusive experiences on social platforms. To reduce risks, the interview attempted to focus on a single abuse instance experienced by each participant. Participants were asked several times during the interview if they were comfortable discussing potentially sensitive topics and were told they could skip any question.

\subsection{Limitations}

The survey received responses from only three transgender and one agender respondents. While they reported exposure to more types of abuse than men and women, these numbers cannot provide statistically significant results. Study findings also cannot be generalized to other participant backgrounds or similar populations outside the US. However, the study reveals that educated, racially diverse young users continue to experience impactful abuse on diverse platforms, and they are able to offer actionable insights and thoughtful recommendations for improving platform abuse defenses.

\section{Findings}
\label{sec:findings:interview}

The survey received 247 responses. We discarded responses from 17 participants who did not have accounts on social platforms, or who answered the survey in less than one minute. Table~\ref{tab:survey:demographics} shows the demographics of the 230 survey respondents. Table~\ref{tab:interview:demographics} shows information about the interview participants. Survey participants are labeled SP1 to SP230, and interview participants are labeled P1 to P15.

\subsection{Abuse Experiences}
\label{sec:findings:interview:experience}

Figure~\ref{fig:socialnetsAll} shows the top three most used social platforms among men and women survey participants (174 men, 47 women). Instagram, Whatsapp, and Twitter/X are the most popular platforms. A Chi-square test reveals a statistically significant association between gender and platform preferences ($\chi^2 = 22.54$, $p < .01$). Notably, Reddit is preferred by more men ($2.13$\% W vs. $34.48$\% M), and LinkedIn is preferred by more women ($19.15$\% W vs. $2.87$\% M) in our sample. A Chi-square test shows, however, no significant association between race and ethnicity and social platform preferences.

\begin{figure*}
\centering
\subfigure[]
{\label{fig:abuse:types}{\includegraphics[width=0.85\columnwidth]{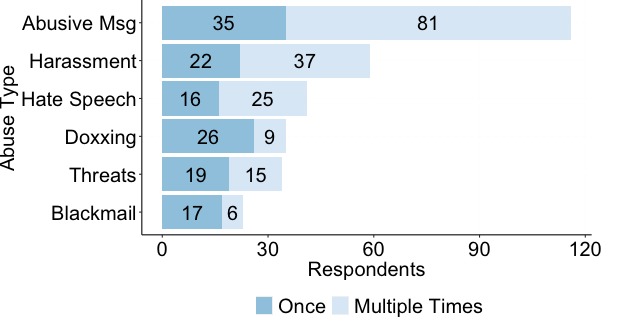}}}
\hspace{25pt}
\subfigure[]
{\label{fig:abuse:impact}{\includegraphics[width=0.85\columnwidth]{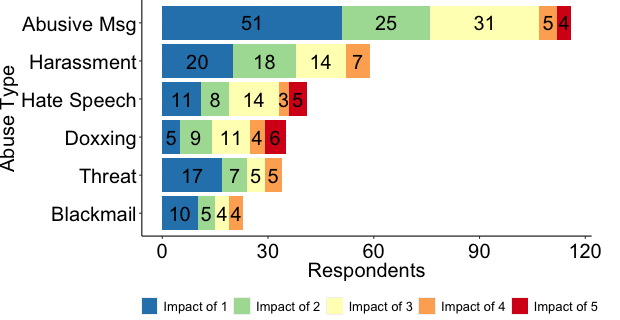}}}
\vspace{-11pt}
\caption{(a) Frequency of participant exposure to different types of abuse through social platforms. Many participants experienced abuse multiple times.
(b) Distribution of abuse impact scores reported by survey respondents for specific instances of abuse they experienced. Respondents reported a total of 43 experienced abuse instances with a score of at least 4.} 
\vspace{-14pt}
\end{figure*}

\noindent
{\bf Types of Experienced Abuse}.
Figure~\ref{fig:sn:mostabuse} shows the platforms on which survey respondents received most abusive messages. Perhaps because people spend more time on popular platforms, they also tend to host more abusive messages. Instagram, the most popular social platform among respondents, was also the one where respondents experienced the most abuse. \newmaterial{While Instagram had the highest number of users (158) and the largest number of abuse reports (40), its normalized abuse report rate, i.e., percentage of users who reported abuse, was (25.31\%). In contrast, Discord had the highest abuse report rate (82.60\%), followed by Twitter (40.27\%), and Facebook (28.26\%).}

Several respondents experienced abuse in online games, even though they did not consider them to be among their top-3 social platforms.

Chi-square tests reveal a significant association between gender and the types of abuse experienced ($\chi^2 = 15.75$, $p < .01$)
, but no significant association between race and experienced abuse types ($\chi^2 = 14.99$, $p = .77$).

Interview participants vividly recalled instances of high-impact abuse. Several participants experienced hate speech based on race, religion, sexual orientation, and politics. In particular, participant P7 experienced political hate speech that escalated into real-life confrontations that involved more students, teachers, and the police. A few participants experienced blackmail e.g.,

\blockquote{
``{\it I received unwanted messages, including sexual images and calls from an unknown person who requested photos, asked me to purchase Bitcoin, and threatened me. The language was intimidating.}'' (P5)}

Participant P5, who experienced this sextortion attack on WhatsApp, further speculated that it was due to a compromise of his personal information after signing up for a free service. Several participants reported threats of physical violence. For instance, participant P14, who experienced abuse in Instagram, related about an argument with an old friend from elementary school that degenerated into threats ``{\it she got mad at me and made threats. She claimed that her boyfriend, who is apparently in a gang, would come to my house because they knew my address, and shoot up my house}'' (P14). For P11, the threats she received through WhatsApp, Snapchat, and Instagram were suggestive of stalking, e.g., ``{\it he made threats like he would travel to [US state of participant], come to my house, find me at my school, and cause harm because I wasn't responding to his messages}'' (P11). This also reveals that abusers exploit victim activities on multiple platforms, including their sharing of sensitive information, e.g., location of school and home, on those platforms. Table~\ref{tab:interview:demographics} further provides the types of abuse experienced by each interview participant.

Figure~\ref{fig:abuse:types} further shows the numbers of survey participants who experienced abuse once vs. multiple times. A majority of respondents who experienced abuse experienced it multiple times. Similarly, in the interviews, a majority of participants reported repeated attempts of contact and abuse from the same person. Some repeat abuse was perpetrated from the same platform, like the case of participant P5, who was blackmailed to prevent doxxing in WhatsApp, ``{\it he contacted me seven times from different accounts with the same threats. I had to block several numbers to stop the threat}''. This reveals that abusers are able to create fake sockpuppet accounts, and victims continue to engage with such accounts. Some participants were abused through multiple platforms, often simultaneously, 
e.g., ``{\it he continued to contact me on WhatsApp, Snapchat, and Instagram, even making phone calls}'' (P11). This shows that abusers leverage multiple platforms where victims are active to amplify their attacks. We later discuss how this reveals the need for victims to be able to link accounts controlled by the same abuser.

\noindent
{\bf Impact of Experienced Abuse}.
The survey asked participants to provide a score, an integer between 1 and 5 (see $\S$~\ref{sec:methods:survey}) for the impact of specific instances of abuse they experienced. Figure~\ref{fig:sn:mostabuse} shows the distribution of impact scores per social platform where abusive messages were received. Spearman's correlation coefficient shows a moderate positive relationship between the popularity of platforms among the respondents and the number of respondents who experienced abusive messages on them ($rs$ = $0.515$). Survey respondents experienced abuse with an impact exceeding 3 only on Instagram, Twitter/X, Discord, and Facebook.

Figure~\ref{fig:abuse:impact} shows the distribution of impact scores per reported type of abuse. While no survey respondent ranked blackmail, threat, or harassment with a score of 5, overall, they reported 43 abuse instances with an impact of at least 4. 
This is consistent with earlier reports of marginalized individuals experiencing higher exposure to online abuse.
The three transgender and one agender survey respondents reported exposure to more types of abuse ($M = 4.0$, $SD = 1.41$) than men ($M = 2.23$, $SD = 1.32$) and women ($M = 1.8$, $SD = 1.01$). This is consistent with earlier reports of marginalized individuals experiencing higher exposure to online abuse~\cite{ADL2022,ADL2021Gaming}. A Mann-Whitney U test ($U = 1332$, $p = .16$) revealed no significant difference between the number of types of abuse experienced by women ($M = 1.8$, $SD = 1.01$) and men (M = $2.23$, $SD = 1.32$). However, a Mann-Whitney U test ($U = 1307$, $p < .05$) revealed that women experienced abuse with significantly higher impact ($M = 2.39$, $SD = 1.08$) than men ($M = 1.83$, $SD = 1.00$). A higher percentage of women in our study experienced hate speech than men ($25.53$\% W vs. $13.79$\% M). Surprisingly, however, a higher percentage of men reported exposure to blackmail ($2.13$\% W vs. $10.92$\% M), threats ($4.26$\% W vs. $16.67$\% M), and doxxing abuse ($4.26$\% W vs. $17.82$\% M) than women.

A non-parametric Kruskal-Wallis test further revealed a significant difference in abuse impact among respondents based on their race and ethnicity ($KW = 22.34$, $p < .001$). Subsequent Mann-Whitney U tests with Bonferroni correction (to mitigate Type I errors due to multiple comparisons) revealed statistically significant differences between several groups. For readability, we use abridged terms for the studied race and ethnicity groups. First, White respondents experienced lower impact abuse ($M = 1.58$, $SD = 0.91$) than (1) Asian respondents ($M = 2.30$, $SD = 1.00$) with a U statistic = $1372.0$ and p-value $< 0.001$, (2) Black respondents ($M = 2.57$, $SD = 1.31$), with U statistic = $663.0$, p-value $< 0.01$, and (3) Hispanic non-White respondents ($M = 2.35$, $SD = 1.21$), with U statistic of $1930.5$ and p-value $< 0.01$. Further, Hispanic White respondents experienced lower impact abuse ($M = 1.81$, $SD = 0.90$) than (1) Hispanic non-White respondents with U statistic of $4141.0$ and p-value $< 0.01$, and (2) Asian respondents with U statistic of $2941.0$ and p $< 0.01$.

Several interview participants discussed abuse incidents with an impact of at least 4. For instance, P4 discussed the long-term effects of harassment experienced through Snapchat, e.g., ``{\it it is deeply ingrained in my memory. Constantly thinking about it surely took a toll on my well-being}'' (P4). Most interview participants continue to feel anxiety, irrespective of impact scores, e.g., ``{\it I already struggle with severe anxiety, and this situation only exacerbated it}'' (P14). Withdrawal from social interactions and trust issues were also common among participants, e.g.,

\blockquote{
``{\it [The abuse] made me withdraw from being around people and made me more reserved. I sought professional help for my anxiety by seeing a therapist and a psychiatrist. I was prescribed medication.} (P11)} 

This reveals that even highly educated young adults experience diverse, high-impact abuse on social platforms, with long-term effects. Coupled with the finding that social platforms can exacerbate trauma~\cite{SMANS23}, it suggests the need for a new approach to discourage abuse and better support victims.

\subsection{Preferences of Platform Abuse Response Affordances}
\label{sec:findings:interview:defenses}

\begin{figure*}[t]
\centering
\subfigure[]
{\label{fig:defenses2}{\includegraphics[width=0.85\columnwidth]{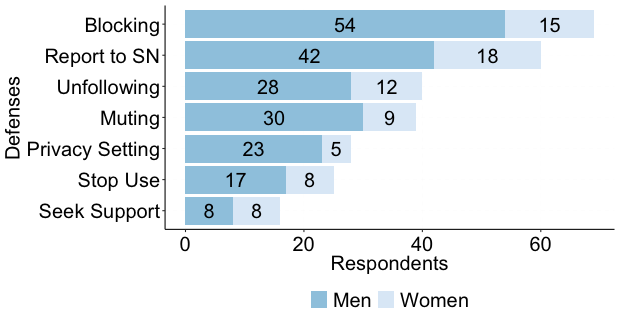}}}
\hspace{25pt}
\subfigure[]
{\label{fig:changes}{\includegraphics[width=0.85\columnwidth]{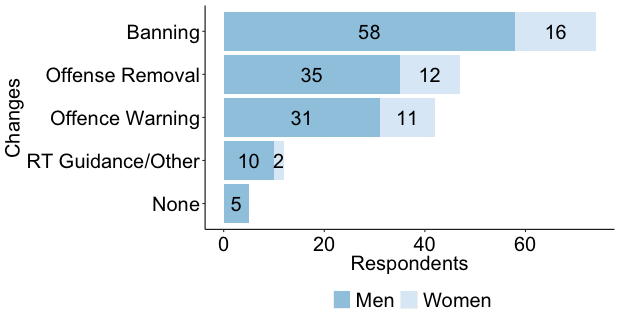}}}
\vspace{-7pt}
\caption{(a) Per-gender distribution of defense strategies adopted by survey respondents after exposure to abuse. Chi-square tests reveal a statistically significant correlation between gender and adopted defenses but not between adopted defenses and race or the impact of experienced abuse.
(b) Per-gender preferences of social platform responses to abuse. Chi-squared tests show no statistically significant relationship between survey respondent suggestions and either gender, race, or impact of experienced abuse.}
\vspace{-14pt}
\end{figure*}

Figure~\ref{fig:defenses2} shows the distribution of abuse defenses employed by the 116 survey respondents who experienced abuse. A Chi-square test shows no statistically significant correlation between participant race and ethnicity and the choice of defenses ($\chi^2 = 7.20$, $p = .97$). However, a Chi-square test reveals a statistically significant correlation between gender (men vs. women only) and the choice of defense strategies in response to abusive messages ($\chi^2 = 12.77$, $p < 0.05$). While similar percentages of men and women adopted blocking ($65.22$\% W vs. $62.07$\% M), muting ($39.13$\% W vs. $34.48$\% M), and changing privacy settings ($21.74$\% W vs. $26.44$\% M), a significantly higher percentage of women 
%
unfollow abusers ($52.17$\% W vs. $32.18$\% M) and report abusers ($73.91$\% W vs. $48.28$\% M). In addition, a Chi-square test reveals statistically significant differences between preferences for responses to abuse based on the impact of abuse experienced ($\chi^s$ = $19.54$, $p$ = $0.012$). In particular, a higher percentage of respondents who experienced high-impact abuse (score of 4 or 5) prefer each of the investigated abuse responses when compared to the respondents who experienced low-impact abuse (score of 1 or 2). We detail this in the following, along with insights we collected from interview participants.

\noindent
{\bf Processing Social Contacts}.
Most of the 116 survey respondents either blocked (73), unfollowed (45), or muted (44) abusers. Recent work also reported higher popularity of the blocking defense~\cite{AFTC24}. However, we find that $73.68$\% (14/19) of respondents who experienced high-impact abuse blocked abusers, which significantly exceeds the $53.08$\% (43/81) among those who experienced low-impact abuse.

Most of the interview participants reported blocking or unfollowing abusers. For instance, to address the high-impact threats and stalking experienced on multiple platforms, participant P11 observes that ``{\it blocking the abuser gave me a sense of security knowing that they no longer have access to view my posts on Instagram}''. However, for participants whose social network is infiltrated by multiple accounts controlled by the same abuser, a single defensive action is insufficient. Instead, a few participants further revisited their online social network after the abuse to remove weak-tie contacts. This includes participant P3, whose account was impersonated on Facebook,

\blockquote{
``{\it I went from having approximately 500 friends to just 90. I removed many people I only knew through social media but hadn't met in person.}'' (P3)}

The participant suspected that only someone who was part of his social network had access to the information required to impersonate his account. His inability to identify the abuser led to a generalized pruning of his social network. The significant impact that abuse can have on the size of the social network of victims, should provide incentives to platform operators to improve their responses to abuse. Participants like P7, who experienced hate speech of moderate impact on Facebook, also applied such defenses proactively before the abuse. e.g., ``{\it I have frequently used the block and unfollow features on social media, not because people directly threatened me, but because their photos or opinions were too controversial}''. While weak-tie contacts~\cite{G83} help maintain and increase social capital~\cite{ESL07, BKM11}, removing those with which users are not comfortable can help reduce the attack surface for future abuse. However, participant P7's experience reveals the difficulty of avoiding abuse even when aware of the dangers of social networking and when diligently applying platform defenses.

For several interview participants, the experienced abuse also made them more conservative when establishing new social contacts. This includes changes to invitation strategies, e.g., ``{\it I only follow friends that I've known for a long time or people that I know in person}'' (P3), and to strategies to process new contact requests, e.g., ``{\it after blocking the abuser accounts, I continued to block and unfollow any new accounts that tried to contact me}'' (P11). Some explained that this is because they suspect that new contact requests may originate from abusers they previously blocked, 

\blockquote{
``{\it I received messages from a different account using a random person's picture, that continued with threats and attempts to intimidate me. She shouldn't have been able to create another account and reach out to me while being blocked.}'' (P14)}

This reveals concerns among participants who experienced high-impact abuse, with the ease of creating fake accounts on social platforms and the failure of platforms to signal to victims information about other accounts controlled by abusers, particularly among their social networks.

\noindent
{\bf Reporting Abuse to the Platform}.
Sixty-five survey respondents have reported abuse to the social platform. The ratio is higher among those who experienced high-impact abuse ($57.89$\%) than among those who experienced low-impact abuse ($44.44$\%). However, 51 respondents did not report abuse, either because of the low severity of the abuse (20), being used to the abuse (15), not being aware that it is possible to report abuse (9), or inability to do it anonymously (7). Victims may prefer anonymous reporting to avoid perceived escalation with the abuser, which may result in further abuse. However, respondent SP24 also raised concerns about the double-edged sword of reporting abuse, ``{\it you think twice about reporting an account because there is a chance you might also get banned for not responding in a correct way}''. Platform support may help victims avoid responding in kind to abuse.

In the interview, all participants acknowledged that reporting abusers is important. For instance, participant P14, whose experience with threats and stalking on Instagram had the maximum impact of 5, does it ``{\it to help protect others from experiencing similar threats or harassment}''. However, only a few interview participants reported the abuse they discussed, to the social platform. This is notably the case for online gamers who often refrain from reporting abuse since they consider it to be typical or acceptable banter~\cite{BFRMK21}. We note that none of the gamers who reported the abuse received a response from the gaming platform. This was also the case for several abuse reports on other social platforms, for instance, for participant P2, who experienced harassment of impact 4 in Discord, ``{\it they said they would look into it, but I didn't hear anything further after that}'' (P2). Other participants often did not report the abuse because of a history of frustrating experiences with the reporting process. This is consistent with previous findings of the reporting process being inefficient~\cite{W17,ashraf2023stalking}, and reporting tools fail to account for individual experiences~\cite{BDSL17}. Several participants believe this is because platforms are overwhelmed by the large volume of received reports, e.g., ``{\it social platforms receive millions of these requests, they cannot act on them in a timely manner}'' (SP24). This suggests the need for changes to the design of abuse reporting systems, to ensure that social platforms have the resources to process abuse reports.

\noindent
{\bf Changing Privacy Settings}.
Thirty survey respondents have changed their privacy settings following the abuse. A higher percentage of high-abuse impact victims change privacy settings (36.84\% = 7/19) than low-impact abuse victims (17.28\% = 14/81). Similarly, most of the interview participants have changed their privacy settings following the abuse. Kekulluoglu et al.~\cite{KVM22} found that Twitter users tend to opt for protected settings to maintain personal content privacy and avoid interactions with strangers and harassment. However, not all social platforms allow such modifications. For instance, interview participant P6, who experienced high-impact discrimination and hate speech in Slack, was unable to modify privacy settings, since `{\it there is no public and private concept in Slack}''. Changing privacy settings is also not a universal solution to abuse: participant P13, who was harassed in online games, pointed out that ``{\it even if your account is set to private, people can still send you abusive messages}''. Making accounts private cannot stop abuse from existing contacts, who can continue to access the user's account information and communication channels.

\noindent
{\bf Seeking Outside Support}.
Nineteen survey respondents coped with the abuse by seeking outside support. Significantly more victims of high-impact abuse (47.36\% = 9/19) sought support than low-impact abuse victims (3.70\% = 3/81). Some of the interview participants who experienced high-impact abuse relied on their friends and family or sought professional help. For instance, participant P14, who experienced threats and stalking of impact 5, reached out to friends and family, ``{\it I informed both my mom and dad about the situation, and as a safety measure, I stayed at a friend's house for a few days}''. Participant P2, who experienced harassment of impact 4, observed that involving third parties can also complicate the issue, ``{\it the friend ended up getting into an argument with the harasser as well [..] it ultimately didn't do anything to resolve the situation}''. Other interview participants chose not to share information about the abuse with others. Reasons to not share include being used to the abuse, feeling embarrassed about being a victim, and not knowing how to discuss the abuse. In particular, participant P4 who experienced harassment of impact 4 perceives a lack of understanding stemming from generational differences, e.g., ``{\it I feel the older generation may not fully understand the complexities of such incidents happening on social media [..] they might have advised me to stop using social media}'' (P4).

\noindent
{\bf Stopping Platform Use}.
In contrast to P4's unwillingness to stop using social platforms despite experiencing high-impact abuse, 28 survey respondents stopped using platforms where they experienced abuse. Again, a substantially higher percentage of victims of high-impact abuse stop using the platforms (57.89\% = 11/19) when compared to low-impact victims (9.87\% = 8/81). Several interview participants reveal that they stopped using the platforms either for a limited interval (weeks to years) or permanently, which is the case for participant P13 who no longer uses Twitter/X after experiencing harassment. A few interview participants deactivated or closed their accounts. This is the case of participant P11, who experienced threats and stalking of impact 5 across Instagram, WhatsApp, and Snapchat. While this action was a direct result of the abuse experienced, P11 only deactivated the Instagram account,

\blockquote{
``{\it I deactivated my Instagram account due to the ease of creating new accounts and the prevalence of impersonation. This eliminated one of the ways the abuser could communicate with me or see my activities. 
I kept my Snapchat and WhatsApp accounts active because they offer relatively more secure features.}'' (P11)}

Thus, the inability to address abuse using the functionality provided by Instagram is perceived to be related to the ease for abusers to create new accounts. Unlike Instagram, WhatsApp requires a phone number to create an account, while Snapchat's Snap Score feature adds a layer of verification by providing insights into a user's activities.

\subsection{Preferences for Platform Responses to Abuse}
\label{sec:findings:interview:recommendations}

Figure~\ref{fig:changes} shows the preferences of 116 survey participants who reported being exposed to abuse, about the abuse defenses that platforms should implement. Chi-square tests reveal no statistically significant association between preferences and gender ($\chi^2 = 1.8$, $p = .63$), between preferences and race and ethnicity ($\chi^2 = 6.9$, $p = .86$), and between preferences and the impact of abuse experienced ($\chi^2 = 6.4$, $p = .17$).

\noindent
{\bf Removing vs. Labeling Abuse}.
More survey respondents (51) prefer that platforms remove abusive content than to label it (45). While the differences are not statistically significant, we observe that more respondents who experienced high-impact abuse (score of 4 or 5) and low-impact abuse (score of 1 or 2) prefer removing to labeling abuse. Removing abuse is recommended by 52.63\% (10/19) of respondents who experienced high-impact abuse and 34.56\% (28/81) of respondents who experienced low-impact, exceeding the 31.57\% (6/19) of high-impact abuse victims and 25.92\% (21/81) of low-impact abuse victims who recommend labeling abuse.

Most interview participants who experienced high-impact abuse also prefer the removal of abusive content to labeling. Participant P14, who experienced threats and stalking with the highest impact (5) on multiple platforms, observes that leaving the content helps the abuser,

\blockquote{
``{\it Even if they label a post as containing false information or a potential threat, users can still view the content. It would be more effective to simply remove such posts.}'' (P14)}

In previous studies by Im et al.~\cite{ISIGWBAFGGN22}, women who experience higher impact abuse also prefer content removal to labeling such content. However, most interview participants who experienced moderate impact abuse prefer labeling instead of removal, ``{\it it is important to maintain a warning system rather than completely removing flagged content}'' (P12), and discussed problems generated by the removal of detected abuse. For instance, echoing calls for platforms to avoid exclusion defenses~\cite{RCHS23}, participants mentioned that side-effects of abuse removal include the perception of censorship:

\blockquote{
``{\it When it comes to various topics, including politics, social platforms can determine who is allowed to post, leading to the blocking of accounts for reasons unrelated to actual abuse.}'' (P10)}

Many interview participants prefer that social platforms automatically detect abuse, which would spare them the effort to report it and convince the operators that the content is abusive. For instance, P9, who experienced threats of impact 4 on Facebook, observes that ``{\it social network platforms could employ algorithms or AI to predict and identify users who engage in abusive language or hateful behavior and take appropriate action}''. Social platforms can indeed leverage the extensive literature work on hate speech detection, e.g.,~\cite{DWMW17}. Participant P4, who experienced harassment of impact 4 through Snapchat, mentioned that benefits of automation include improved reaction times, ``{\it posting inappropriate, disturbing content is immediately blocked, and messages are deleted even before the recipient reads them}''. We observe the perceived benefit of victims not being exposed to abusive content. Participant P6, who experienced hate speech in Slack, also mentioned that receiving confirmation is another benefit for platforms detecting abuse because ``{\it sometimes you don't know yourself whether you are abused or not}''. This suggests the need for platforms to provide formal confirmation of abuse to victims.

Participants, however, also discussed challenges in the automated detection of abuse. Given the evolving nature of online threats~\cite{APP19}, participants raised the issue of incorrect classification. Incorrect labeling of content can occur due to the difficulty of defining abuse due to subjective and cultural differences, e.g., ``{\it what might be considered offensive to Nepali people may not be offensive to Indian people and vice verse}'' (P7). This is consistent with the findings of Jhaver et al.~\cite{JGBG18} that not all users agree on what constitutes harassment. Participant P15, who was harassed on Discord, observes that ``{\it if someone wants to convey a threatening message and knows it might be flagged, they will use coded language}''. This suggests that false negatives in the classification of abuse occur also when abusers find ways to circumvent moderation.

\noindent
{\bf Suspending Abusers}.
The most popular defense, recommended by 78 survey respondents, is suspending abusers from the platform. Abuser account suspension is recommended by a higher percentage of respondents who experienced high-impact abuse (57.89\%) and low-impact abuse (61.72\%) than those who recommended removing or labeling abusive content. Interview participants observe that account suspensions can discourage would-be offenders. For instance, participant P11, who experienced stalking and threats of impact 5 on several platforms, believes that the blocking of abusers by victims is not sufficient, and social platforms need to get involved,

\blockquote{
``{\it I wish social media platforms would actively monitor user accounts and ban those who abuse others. This would discourage individuals from spreading hate or engaging in harmful activities.}'' (P11)}

Rather than suspending abusers, participant P12, who experienced medium-impact harassment on Instagram, suggested adopting a technique implemented by online games that ``{\it have a symbol or marker on a player's name to indicate their rude behavior}''. Participant P12 further recommended that labeled abusers should be restricted to interacting only with other abusers. However, other participants observed that suspended (or abuse-labeled) users can easily create new accounts and continue their abusive behaviors. This is because of the small amount of data required by some platforms for users to create accounts, as pointed out by P1, who was harassed and doxxed on several platforms, ``{\it Tumblr only requires an email confirmation to create an account}''.

This suggests that the anonymity provided by social platforms' failure to detect fake, sockpuppet accounts, reduces the effectiveness of measures taken against abusive users.

Consequently, participants like P8, who experienced hate speech on TikTok, recommend that social platforms generalize account verification, e.g., ``{\it the social network should enforce users to provide their real information}''. This includes platforms providing mechanisms to verify all claims made by users on their profiles. While some apps like WhatsApp require a phone number to create an account, platforms like TikTok or Instagram do not mandate even such data collection at account creation time. Ironically, Douyin, the TikTok version developed for domestic consumption in China, requires users to sign up using their national ID number and a SIM card registration to that number.

\noindent
{\bf Improving Responses to Abuse Reports}.
In addition, several interview participants emphasized the need for platforms to improve their responses to abuse reports. This is consistent with recent reports of insufficient abuse reporting processes in social media~\cite{AFTC24}. For instance, participant P3, who reported an impersonator to Facebook, would like to see an improved attitude toward victims,

\blockquote{
``{\it [Social platform operators] shouldn't act defensively or aggressively towards me. I have noticed that they don't believe me even after I have provided all the necessary information.}'' (P3)}

Participant P9, who received high impact (4) threats on Facebook, would like to receive an acknowledgment or feedback after abuse reports, ``{\it I want the social network to acknowledge and investigate it, and take appropriate actions}'' (P9), along with reducing response times to abuse reports, e.g., ``{\it I expect them to respond promptly when I report an issue}'' (P9). 
This suggests that social platforms need to improve both moderator responses and automated responses to abuse reports, to provide victims with more feedback during the verification process. 
ARI builds on these recommendations to provide automatic certification of abuse and personalized defenses that support victims in their handling of the abusers and preventing future abuse.

\subsection{Perceptions of Privacy and Anonymity}

Interview participants reveal a complex interplay of the privacy and anonymity features provided by social platforms. In many instances of abuse discussed during interviews, the participants were personally acquainted with their abusers, who included friends and colleagues. When not having access to information about the abuser, participants, however, reported increased anxiety due to unpredictability. For instance, participant P8 who experienced hate speech in TikTok:

\blockquote{
``{\it I visited the abuser page, but there was no content, and there was no profile picture either. It was a faceless abuser, and that made me feel very unsettled, paranoid and anxious because it could have been anyone, they could be someone I know.}'' (P8)}

This suggests that not knowing the identity of the abuser amplifies the impact of the abuse, e.g., making some victims experience generalized distrust toward their entire social circle. We observe that some participants also use the social platform to collect information about the abuser. For instance, P14 monitored the abuser's activities, e.g., ``{\it there were moments when I would unblock [the abuser account] just to check if it was still active on the platform}''. Participants also use the social platform to collect evidence of abuse, a vital process for users to prove that they have been victims of abuse. However, privacy and security features of some platforms prevented several interview participants from collecting evidence: 

\blockquote{
``{\it I wasn't able to capture a screenshot because of Snapchat's disappearing message feature. Also, before I could take a screenshot, [the abuser] had already blocked me and vanished from my list.}'' (P4).}

This reveals that some abusers have adopted behaviors to evade retribution for their actions, including leveraging platform features and evolving hit-and-run strategies. An abuser blocking a victim on Snapchat ensures that the victim will no longer be able to access the profile or history of messages received from the abuser. This suggests that even highly educated victims need support from social platforms to preserve and collect evidence about the abuse.

When interview participants use social platforms to connect with high-value contacts, they typically open accounts using their real identity in order to establish trust with the contacts. When using platforms to connect with weak-tie contacts~\cite{G83}, participants tend to prefer pseudonyms or nicknames instead of real names. For instance, all the interview participants who play online games use pseudonymous accounts on gaming platforms. For participant P12, who is an online gamer who has experienced no abuse on gaming platforms, one reason for this is the perceived toxicity of gaming platform interactions and the associated risks of revealing their identity, e.g.,
\blockquote{
``{\it I prefer to prevent other players from tracking my profile or trying to contact me outside of the game. I want to ensure that the interactions stay within the confines of the game.}'' (P12)}
The anonymity provided by nicknames is also used by participants in other social platforms to sandbox their activities to certain accounts. Participant P9 who learned from high-impact threats experienced on Facebook, explained that ``{\it it's important not to use real names when creating online profiles because it makes it easier for others to identify you}''. We observe, therefore, that anonymity is a double-edged sword that absolves abusers of accountability for their actions but can also protect skilled users from abuse.

\section{Victim-Centric Abuse Defenses}
\label{sec:discussion}

\subsection{Summary of Study Findings}

Several social platforms emerge to be more conducive toward abuse. They include Snapchat, with its message disappearance feature, and platforms like Instagram and WhatsApp, perhaps due to their popularity. While some participants master the diverse arsenal provided by social platforms to defend against abuse, some abusers also skillfully leverage the anonymity provided by most platforms to avoid punishment and amplify victim anxiety. Participants also revealed forms of abuse cross-pollination on multiple social platforms, where abusers leverage the range of existing platforms to both collect background information about non-anonymous victims and amplify their abuse by directing their attacks from multiple sources, particularly after being blocked by victims.

Previous work has shown that women and minority groups are more vulnerable than men to online harassment~\cite{SBDDGWKTA23, ISIGWBAFGGN22, FF22, UCW21}. Our survey paints a more complex picture. First, we find that Asian, Black, and Hispanic non-White respondents experienced significantly higher impact abuse than White and Hispanic White respondents. However, we found no statistically significant correlation between participant race and ethnicity and the choice of defenses. In addition, while a higher percentage of women experienced hate speech than men, a higher percentage of men reported exposure to blackmail, threats, and doxxing abuse than women. However, women experienced abuse with a statistically significantly higher impact than men. Gender also has a significant association with survey respondent choices of protective measures following abuse, where a significantly higher percentage of women change what they post, unfollow abusers, and report abusers. Differences in the impact of abuse experienced based on participant background and the different types of abuse prevalent on the diverse platforms available suggest the need for more personalized defenses.

Our study provides complementary evidence to existing literature knowledge that many users are skeptical about the fairness of resolutions following online abuse reports~\cite{SSHCS21, TAS22, sambasivan2019they, wolak2016sextortion}. In particular, a majority of the survey respondents who did not report abuse were either being used to the abuse or were not aware that it was possible to report it. Interview participants recommend that social platforms provide timely responses to their abuse reports and provide evidence of incorporating their feedback into platform operations. This suggests the need for social platforms to initiate personalized defenses and interventions when users report abuse.

In addition, participant discontent with platform abuse responses sometimes stems from inconsistencies between the defenses implemented by different platforms. Some participants describe suggestions for platforms where they experienced abuse by drawing parallels to features available in other platforms where perhaps they experienced less abuse. This suggests the need for more consistent features offered by different platforms. This is not uncommon. For instance, some of X's recent changes were adopted by other platforms, e.g., Meta adopting paid account verification in Facebook and Instagram~\cite{R23}, and Reddit charging for API accesses~\cite{L23}.

Study participants revealed a subtle and conflicting interplay between privacy, anonymity, and attribution of abuse on social platforms. While most study participants changed their privacy settings following abuse, they also revealed that doing so does not protect them from all types of abuse. This is because users with private accounts continue to be vulnerable to abuse from their existing social contacts, who can still access their profiles, posts, and direct communication channels. In addition, while anonymity helps users protect their identity from weak tie contacts and sandbox their activities on different accounts or platforms, it also amplifies the impact of abuse experienced by non-anonymous users. This suggests the need to re-design social platform collection and use of user-identifying information, to simultaneously enable well-behaved pseudonymous interactions and provide vital defenses to abuse victims.

\subsection{Abuse Defense Design Implications}

We build on the above findings, viewed through the lens of constraints experienced by social platform operators, to introduce several design requirements for abuse response systems:

\begin{compactitem}

\item
{\bf Unified, transparent abuse responses}.
Social platforms should provide consistent processes for users to report abuse and to process and respond to abuse reports. The verification process needs to be transparent for the user reporting the abuse.

\item
{\bf Document abuse}. Platforms should provide mechanisms for victims to capture, persist, and later retrieve evidence of abusive interactions.

\item
{\bf Alternatives to permanent suspensions}.
Social platforms should provide alternatives to suspensions that disincentivize abuse.

\item
{\bf Prevent abuse of abuse-reporting systems}.
Malicious users should find it difficult to abuse platform abuse-reporting systems, e.g., through denial of service attacks that seek to cripple platform responses to legitimate reports, or by framing and victimizing other users.

\item
{\bf Sustainable abuse defenses}.
The processes to identify and respond to abuse should not place undue burden on platform operators.

\end{compactitem}

\subsection{Blueprint for Abuse Responses and Interventions}

We build on these requirements to introduce ARI, a blueprint for social platforms to provide a unified, personalized defense interface to abuse victims.

\noindent
{\bf Abuse Reports}.
Users who report abuse need to reference the specific abusive posts. However, abusive posts may be removed, e.g., by abusers who seek to erase evidence. To satisfy the abuse documentation requirement, ARI-compliant platforms preserve posts for a set interval, e.g., 6 months, after being erased. Erased posts are no longer directly accessible to users unless they reference them in abuse reports.

In order to sustainably validate abuse reports, ARI builds on two insights. First, the cost of defenses should be covered by abusers. Second, social platform users can help validate reported abuse. We detail these next.

\noindent
{\bf Abusers Cover Defense Costs}.
To discourage abusive reporting behaviors, ARI-compliant platforms can impose a fee to be paid by the user reporting the abuse. Exemptions may be provided to special status users, e.g., minors, who have a clean record of using the reporting system. To avoid imposing additional financial burden on disadvantaged victims, the fee can be kept in escrow during the verification process. If the reported abuse is validated (see next paragraph), the fee is refunded to the victim; the platform can then charge the penalty fee to the abuser. The collected fees, which may be paid through online currencies like Facebook's Viewpoints~\cite{Viewpoints}, can be shared by the platform with the reporting abuse victims~\cite{ISIGWBAFGGN22}. Payments may be applied on a per-abuse instance and per-victim basis and may increase with repeat offenses.

This approach charges the funds required for the verification process either to abusive reporters or to verified abusers. ARI-compliant platforms can prevent abusers from using their accounts until they provide the payment. This enables the collection of additional abuser PII, which can be used to identify and link pseudonymous accounts to known abusers.

\noindent
{\bf Crowdsourcing Abuse Verification}.
The platform can use parts of the collected fee to fund the abuse verification process. To reduce the burden of the abuse report verification process, and address issues raised in existing literature~\cite{sambasivan2019they, wolak2016sextortion, TAS22, SSHCS21} (see $\S$~\ref{sec:related}), ARI-compliant platforms can crowdsource the process to other platform users. More specifically, each abuse report is anonymized and presented to $k$ verifiers. Each verifier decides whether the report contains abuse and scores its perceived impact. The platform aggregates responses, e.g., majority voting or weighted sum.

\noindent
{\bf Abuse Certification}.
Most interview participants believe it is important to receive an acknowledgment from the platform upon reporting abuse. To achieve this, when a user reports abuse, an ARI-compliant platform should ensure the existence of at least one post documenting the abuse. The platform issues the reporting user with a signed acknowledgement of receiving the abuse report. The acknowledgment needs to include a cryptographic hash of the report and a timestamp of receipt.

After the above abuse verification process, if the reported abuse is validated, the platform issues the victim with a signed {\it abuse certificate} that includes the type of abuse, a description of the abuse, and the time when it took place.

\section{Conclusions}

Through a survey and semi-structured interviews, we studied the abuse experienced by young, educated minority users on social platforms, their proactive and reactive use of platform affordances to self-protect, and their perceptions of platform defenses. The work reveals the complex interplay of privacy, anonymity, and abuse attribution issues underlying the design of abuse defense systems. We have built on study findings to propose design goals for abuse defense systems, and to introduce ARI, a blueprint for a defense approach that encourages collaboration between platforms and abuse victims.

\section*{Acknowledgments}

\newmaterial{
This work was supported in part through NSF awards 2321649, 2114911, 2013671, 2244283, and 2055485. We thank Yassel Pena, Ladislau Boloni, Michelle Taub, Kendra Husband, Frideline Bruno, and Nicholas Gallegos for early feedback on the survey}.

\bibliographystyle{IEEEtran}
\bibliography{cyberabuse}


\appendix


\subsection{Survey Questions}
\label{appendix:survey}

\subsubsection{Social Media Engagement Overview}

\begin{itemize}
    \item[S-1.] Do you use social networks or communication apps?
    
    \item[S-2.] Which social platforms do you use?
    
    \item[S-3.] What are your top three most used social platforms?
    
    
    \item[S-4.] What other social platforms do you use on your devices?
    
\end{itemize}

\subsubsection{Abuse Exposure}

\begin{itemize}
    \item[S-5.] Abuse messages include messages that are inappropriate, offensive, or insulting.
                Inappropriate: Not suitable or proper within the context or setting in which it occurs.
                Offensive: Causing someone to feel hurt, upset, or angry through derogatory or demeaning content.
                Insulting: Expressing disrespect or contempt, often in a way that is scathingly hurtful.
    Do you recall receiving an inappropriate offensive or insulting message on a social platform? (b) On what social platform did you receive the most inappropriate offensive or insulting messages?

    \item[S-6.] Threats are declarations of the intention to cause harm or adversity to someone in retribution for something done or not done. Has anyone ever threatened you or your family through social networks or communication apps?

    \item[S-7.] Doxxing is the act of publicly revealing previously private personal information about an individual without their consent, often with malicious intent. Has anyone ever doxxed your private media (e.g., images, videos, or text) through social networks or communication apps?

    \item[S-8.] Blackmail is the action of demanding payment or another benefit from someone by threatening to reveal compromising or damaging information about them. Has anyone ever blackmailed you on a social network or communication app?

    \item[S-9.] Harassment involves creating an intimidating, hostile, or offensive environment through unwelcome and uninvited comments. Have you ever experienced any kind of harassment through your social network accounts or communication apps?

    \item[S-10.] Hate speech is any form of communication that disparages a person or a group based on characteristics that include race, color, ethnicity, gender, sexual orientation, nationality, or religion. Have you ever felt discriminated against on social platforms? For instance, because of your race, gender, sexuality, age, or hobbies.

\end{itemize}

\subsubsection{Abuse Impact and Response}.
The following questions are asked after each of the above questions that was answered in the affirmative.

\begin{itemize}
    
    \item[S-11.] On a scale of 1-5, how much did this abuse negatively impact you?
    
    \item[S-12.] Which defenses did you implement against the abuse? (a) Blocking, (b) Changing privacy settings, (c) Changing what you post, (d) Muting, (e) Unfollowing, (f) Other (Please provide details)
    
    \item[S-13.] What changes would you like to be implemented by social platforms to reduce such abuse? (a) Identification and removal of offensive messages, (b) Identification and warning labels on offensive messages, (c) Banning of repeat offenders from the network, (d) Other (Please provide details)

\end{itemize}

\subsubsection{Demographic Information}

\begin{itemize}
    \item[S-14.] With which racial and ethnic group(s) do you identify?
    
    \item[S-15.] How do you describe your gender identity?
    
    \item[S-16.] How old are you?
\end{itemize}

\subsubsection{Additional Information}

\begin{itemize}
    \item[S-17] We plan to conduct follow-up interviews with persons interested in providing additional information. If you want to participate in a follow-up interview, please enter your email address below.
\end{itemize}

\vspace{15pt}
\subsection{Interview Guide}
\label{appendix:interview}

\subsubsection{ Abuse Impact and Responses}.

\begin{itemize}
    \item[I-1.] Do you remember experiencing abuse in any social platform, that had a significant impact on your life? (a) Are you comfortable discussing such an incident?
    \item[I-2.] [Only if answers to I-1 were both affirmative] Can you tell me about the incident, including the context and the social platform where it occurred?
    \item[I-3.] Did you know the abuser? (a) Can you tell me about your relationship?
    \item[I-4.] Do you know how the abuser obtained your personal information?
    \item[I-5.] What was the impact of the incident on your life?
    \item[I-6.] Did you report the incident or the abuser to the social network?
    \item[I-7.] Did the abuser attempt to contact or pursue you AGAIN at a later time? How did you respond?
    \item[I-8.] Did you take any steps to protect against this type of abuse following the incident? 
    \item[I-9.] Were any of these actions effective for you? Why do you think this is the case?
    \item[I-10.] Looking back at this event, is there anything that you would like to have done differently?
    \item[I-11.] Did you seek support from anyone during the period of abuse? If so, (a) Who did you reach out to? (b) Did it help?
\end{itemize}

\subsubsection{Perceptions of Social Platforms}.

\begin{itemize}
    
    \item[I-12.] Do you think that the social network has done enough to prevent this abuse? (a) What do you think they could have done better?
    \item[I-13.] What kind of support would you like to have had to better handle this abuse incident?
    \item[I-14.] Did the social network provide enough support for you to cope with this abuse once it happened? If yes, (a) What kind of support did it provide? If not, (b) How do you feel about this lack of support?
    \item[I-15.] Do you feel that the social network provides enough tools for you to protect yourself from further abuse?
    \item[I-16.] Is there anything that social networks could do better? If you were working for the social platform, what would you change in their abuse response policies and tools?

\end{itemize}

\end{document}